\documentclass[aps,twocolumn,showpacs,preprintnumbers,amsmath,amssymb]{revtex4}
\usepackage{latexsym}
\usepackage{psfig,graphicx}
\usepackage{bm}% bold math

\newcommand{\bea}{\begin{eqnarray}}
\newcommand{\eea}{\end{eqnarray}}
\newcommand{\be}{\begin{equation}}
\newcommand{\ee}{\end{equation}}
\newcommand{\pkt}{\; .}

\newcommand{\tr}{{\rm Tr}}

\newcommand{\calm}{{\cal M}}

\newcommand{\cals}{{\cal S}}

\newcommand{\call}{{\cal L}}

\newcommand*{\del}{\partial}

\begin{document}
\preprint{DO-TH-03/07}
\date{\today}
\title{\bf Quantum dynamics of $\Phi^4$ field theory 
 beyond leading order in $1+1$ dimensions}
\author{J\"urgen Baacke}
\email{baacke@physik.uni-dortmund.de}
\author{Andreas Heinen}
\email{andreas.heinen@uni-dortmund.de}
\affiliation{Institut f\"ur Physik, Universit\"at Dortmund,
D - 44221 Dortmund , Germany}
\begin{abstract}
We consider the out-of-equilibrium evolution of a classical 
condensate field $\phi=\left<\Phi\right>$ and its quantum fluctuations 
for a $\Phi^4$ model 
in $1+1$ dimensions with a double well potential. We use the two-particle 
point-irreducible (2PPI) formalism in the two-loop approximation. 
We compare our results to those obtained in the Hartree approximation,
in the bare vertex approximation (BVA) and in the two-particle 
irreducible next-to-leading order large-N (2PI-1/N) approach,
with thermal initial conditions. In the 2PPI scheme we find that the 
system tends to the symmetric configuration at late times,
as expected in the absence of spontaneous symmetry breaking.

\end{abstract}
\maketitle

\section{\label{sec:intro}Introduction}
\setcounter{equation}{0}
In recent years there was considerable activity in extending the
study of quantum field theory out of equilibrium 
beyond the leading order one-loop, large-$N$ and Hartree 
approximations \cite{Berges:2000ur, Berges:2001fi,
Blagoev:2001ze,  Aarts:2002dj, Cooper:2002ze,Berges:2002cz, Cooper:2002qd,
Baacke:2002ee}.

The simple $\Phi^4$ quantum field theory serves as a toy model 
for more realistic models with more degrees of freedom. Such models 
are e.g. candidates for a description of the inflationary phase 
of the early universe.

Recent numerical simulations 
\cite{Cooper:2002qd, Baacke:2002ee, Mihaila:2003mh} 
within different approximation schemes beyond leading order have lead to
contradictory results for the phase structure of O(1) and O($N$) $\Phi^4$ 
models in $1+1$ dimensions. From finite temperature field theory it is known 
\cite{Griffiths}, 
that in $1+1$ dimensions there should be no spontaneous symmetry breaking 
except at zero temperature; hence the absence of spontaneous symmetry breaking 
is expected in nonequilibrium quantum field theory as well. 
While the Hartree approximation gives a first order phase transition, 
the situation is different for various next-to-leading order approximation 
schemes.

Cooper et al. \cite{Cooper:2002qd}
 have studied in detail 
the symmetry structure of the O(1) model in $1+1$ dimensions 
and the equilibration of
the one- and two point correlation functions within
the bare vertex approximation (BVA) and the two-particle irreducible 
next-to-leading order large-N (2PI-1/N) scheme. They found a 
second-order phase transition in the BVA and no spontaneous symmetry breaking 
in the 2PI-1/N approximation. Both approximation schemes are based on the 
so called Cornwall-Jackiw-Tomboulis 
(CJT) effective action \cite{Cornwall:1974vz}. In the CJT formalism 
the effective action is expanded in terms of two-particle irreducible (2PI) 
diagrams. The self energy fulfills a Schwinger-Dyson equation and 
resums this 2PI diagrams. A related approximation proposed by 
Verschelde and Coppens \cite{Verschelde:1992bs, Coppens:1993zc} 
is the two-particle point-irreducible (2PPI) formalism, where only 
local contributions to the self energy are considered. The one-loop 
2PPI approximation represents the Hartree approximation. We have studied, in 
Ref.~\cite{Baacke:2002ee}, the $O(1)$ model within the two-loop 2PPI 
approximation and, with the parameters and initial conditions
chosen there, we found no spontaneous symmetry breaking.  
In order to have a direct comparison with the results
of Cooper et al. we present here  numerical simulations using  their
 parameters and initial conditions. Besides the question of
spontaneous symmetry breaking it is interesting to compare the general 
features in the temporal evolution of all three approaches.

The plan of this paper is as follows: In section~\ref{sec:model} we briefly
summarize the model formulated in \cite{Baacke:2002ee} and its extension
to the case of an additional thermal initial state of quanta. In
section~\ref{sec:results} we present the results of the numerical
calculations. We end with some conclusions in section~\ref{sec:conclusions}.

\section{\label{sec:model}The model}
\setcounter{equation}{0}
The Lagrange density for the $\Phi^4$ quantum field theory is defined by
\bea
\call=\frac{1}{2}\del_\mu\Phi\del^\mu\Phi-\frac{1}{2}\mu^2\Phi^2
-\frac{\lambda}{4!}\Phi^4 \ .
\eea
Here we will consider the case of a double well potential with $\mu^2<0$. 
Within the 2PPI formalism the inverse Green's function 
$G^{-1}(x,x')=i [\Box + \calm^2(x)]\delta(x-x')$
remains local and is parametrized by a variational mass
$\calm^2(t)$. The 2PPI effective action in terms of the variational
parameters $\phi(t)$ and $\calm^2(t)$ denotes \cite{Baacke:2002ee}
\bea \nonumber
\Gamma[\phi,\calm^2]&=&
\int d^2x \left[\frac{1}{2} \del_\mu \phi(x)\,
  \del^\mu \phi(x) - \frac{1}{2}\calm^2(x)\phi^2(x)\right.
\nonumber\\
&&\hspace{1cm}+\left.\frac{\lambda}{12}\phi^4(x)+ \frac{1}{2\lambda}\left(\calm^2(x)-\mu^2\right)^2\right]
\nonumber\\
\label{eq:gammaphicalm} 
&& + \Gamma^{\rm 2PPI}[\phi,\calm^2]
\pkt
\eea 
The term $\Gamma^\mathrm{2PPI}[\phi,\calm^2]$ contains all 2PPI 
vacuum diagrams, i.e. all graphs that do not decay if 
two lines meeting at the same point are cut.

By variation of the effective action $\Gamma[\phi,\calm^2]$ with respect to 
$\calm^2$ 
one easily checks that $\calm^2(x)$ satisfies the gap or Schwinger-Dyson
equation 
\be \label{eq:calmdef}
\calm^2(x)=\mu^2+\frac{\lambda}{2}\phi^2(x) +\frac{\lambda}{2}
\Delta(x)\ ,
\ee
where the insertion $\Delta(x)$ has been defined as
\be
\Delta(x)=-2 
\frac{\delta \Gamma^{\rm 2PPI}[\phi,\calm^2]}{\delta \calm^2(x)} \ .
\ee
This form of the effective action in terms of $\phi$ and $\calm^2$ is
analogous to the one with $\phi$ and $\Delta$ proposed by 
Verschelde and Coppens \cite{Verschelde:1992bs, Coppens:1993zc}.  

\begin{figure}
\begin{center}
\vspace{0.5cm}
(a)\includegraphics[scale=0.4]{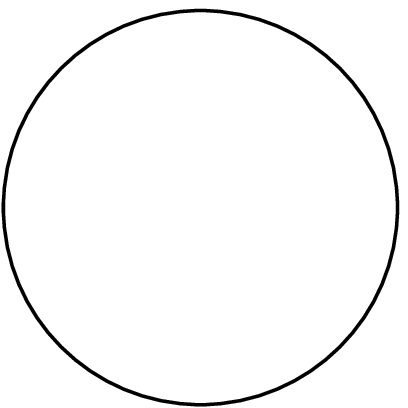}
\hspace{1cm}(b)\includegraphics[scale=0.4]{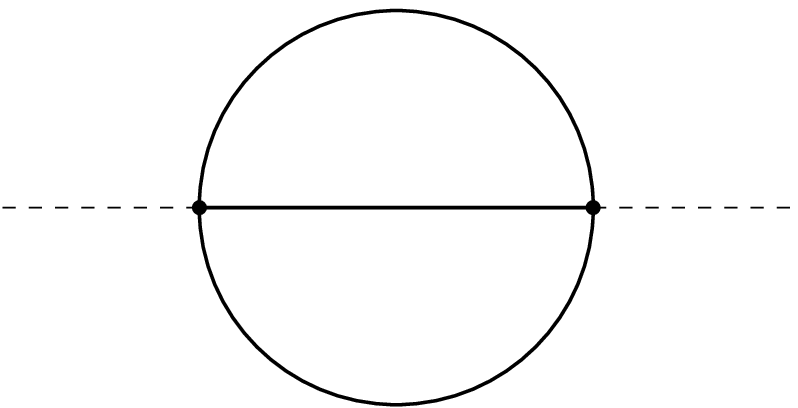}
\end{center}
%{\bf Fig. 1} 
\caption{\label{fig:vacgraphs}We display the leading diagrams in 
the 2PPI action $\Gamma^{{\rm 2PPI}}$: (a) the bubble diagram;
(b) the sunset diagram; solid lines: internal
propagators; dashed lines: external fields $\phi$.}
\end{figure}

By truncation of the infinite series of 2PPI graphs in
$\Gamma^\mathrm{2PPI}$ one obtains an approximation for the effective
action. The two graphs depicted in Fig.~\ref{fig:vacgraphs} 
represent the relevant
contributions to $\Gamma^{\rm 2PPI}$ in the two-loop approximation,
that will be used here. If only the one-loop bubble graph is included we
refer to it as Hartree approximation.

In the following we will consider only spatially homogeneous fields. Time
integrations are understood to be carried out along a closed time path (CTP)
\cite{Schwinger:1961qe, Keldysh:1964ud, Calzetta:1987ey}.

The initial density matrix of the system 
is characterized by a nonzero value $\phi(0)$ 
for the mean field and a distribution function $n_p$ for the quanta. 
We will assume a Bose-Einstein distributed initial state of quanta
parametrized by 
\bea
n_p=\frac{1}{e^{\omega_p/T_0}-1} \ 
\eea
with $\omega_p=\sqrt{p^2+m_0^2}$, where $m_0^2$ is defined below. 
The Green's function can therefore be written in terms of mode functions as 
\bea
G(t,t';p)&=&\frac{1}{2\omega_p}\left(2n_p+1\right)\left[
f(t,p)f^*(t',p)\Theta(t-t')\right.\nonumber\\
&&\qquad \left.+f(t',p)f^*(t,p)\Theta(t'-t)\right] \ .
\eea
The mode functions $f(t,p)$ satisfy the differential equation
\bea
\ddot{f}(t,p)+\left[p^2+\calm^2(t)\right]f(t,p)=0 \ .
\eea
The initial mass $m_0=\calm(0)$ has to be determined self consistently from
the renormalized equation
\bea
m_0^2=\mu^2+\frac{\lambda}{2}\phi^2(0)
+\delta \mu_\mathrm{fin}^2
+\frac{\lambda}{2}\int\frac{dp}{2\pi}\frac{n_p}{\omega_p}
\ ,
\eea
with $\delta \mu_\mathrm{fin}^2=\frac{\lambda}{8\pi}\ln\frac{\mu^2}{m_0^2}$.
The mode functions at $t=0$ satisfy $f(0,p)=1$ and $\dot{f}(0,p)=-i\omega_p$.

The renormalized equations of motion in the two-loop approximation are 
\bea
0&=&\ddot\phi(t)+\calm^2(t)\phi(t)-\frac{\lambda}{3}\phi^3(t)+\cals(t) 
\label{eq:eqmo-phi}
\\
\calm^2(t)&=&\mu^2+\delta \mu^2_\mathrm{fin}
+\frac{\lambda}{2}\left(\phi^2(t)+\Delta^{(1)}_\mathrm{fin}(t)
+\Delta^{(2)}(t)\right)\label{eq:eqmo-M2}\nonumber \\
\eea
The finite one-loop part $\Delta^{(1)}_\mathrm{fin}$ has been defined as
\bea
\Delta^{(1)}_\mathrm{fin}(t)=
\int\frac{dp}{2\pi2\omega_p}\left[\left(2n_p+1\right)|f(t,p)|^2-1\right] \ .
\eea
The two-loop contributions are given by
\bea
\cals(t)&=&
-i\frac{\lambda^2}{6}
\int_0^t dt'\phi(t')\nonumber \\
&&\times\int\prod_{\ell=1}^3 \left(\frac{dp_\ell}{2\pi}\right)
2\pi \delta\left(\sum_{\ell=1}^3 p_\ell\right)\nonumber \\
&&\times\left[\prod_{\ell=1}^3 G(t,t';p_\ell)
-\prod_{\ell=1}^3 G(t',t;p_\ell)\right] \label{eq:cals}\
\eea
and
\bea
\Delta^{(2)}(t)&=&-\lambda^2
\int_0^{t}{dt'}\phi(t')\int_0^{t'}{dt''}\phi(t'')\nonumber \\
&&
\times\int\prod_{\ell=1}^3\left(\frac{dp_\ell}{2\pi}\right)
2\pi\delta\left(\sum_{\ell=1}^3p_\ell\right)\nonumber
  \\
&&\times\left[G(t,t';p_3)- G(t',t;p_3)\right] \\
&&\times
  \left[G(t',t'';p_1)G(t',t'';p_2)G(t,t'';p_3)\right.\nonumber \\
&&\quad-
\left.G(t'',t';p_1)G(t'',t';p_2)G(t'',t;p_3)\right]\nonumber \pkt
\eea
For the BVA and 2PI-1/N approximation Cooper et al. 
introduce the mass term $\chi(t)$ as an auxiliary field.
Naively, in our scheme this corresponds to
the quantity
\bea
\chi(t)&=&\mu^2+\delta
\mu^2_\mathrm{fin}+\frac{\lambda}{2}
\left[\phi^2(t)+\Delta^{(1)}_\mathrm{fin}(t)\right]\nonumber\\
&=&
\calm^2(t)-\frac{\lambda}{2}\Delta^{(2)}(t) \label{eq:chidef}
\eea
where we have identified $\tr G$ with $\Delta^{(1)}$.
We note, however, that the Green's functions have a different
meaning in their and our schemes; this should be kept in
consideration when comparing the numerical results.

%%%%%%%%%%%%%%%%%%%%%%%%%%%%%%%%%%%%%%%%%%%%%%%%%%%%%%%%%%%%%%%%%%%%%%%%%

\section{\label{sec:results}Results}
The numerical implementation for the solution of the equations of
motion [see Eq.(\ref{eq:eqmo-phi}) and (\ref{eq:eqmo-M2})] 
has been described in detail in Ref.~\cite{Baacke:2002ee}. 
The equations of motion are solved using a Runge-Kutta algorithm 
with a time discretization $\Delta t=0.001$. The energy is conserved 
within five significant digits. We have chosen  $\mu^2=-1$, 
$\phi(0)=0.4$, $T_0=0.1$ and $\lambda$ equal to 
$3.0$ and $21.3$. This choice of parameters corresponds to the one of 
Cooper et al. \cite{Cooper:2002qd}.

\begin{figure}[htbp]
  \centering
(a)\hspace{-0.5cm}\includegraphics[width=8cm]{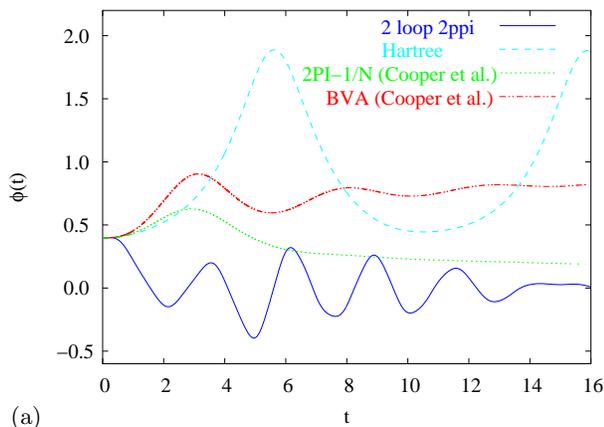}
\par \vspace{0.5cm}
%\qquad 
(b)\hspace{-0.5cm}\includegraphics[width=8cm]{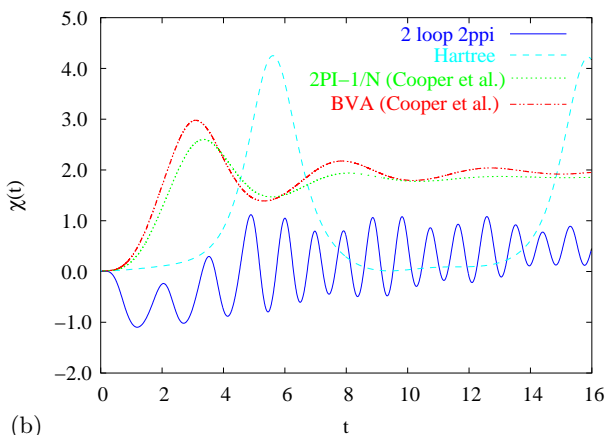}
\caption{\label{fig:phi-l3.0}Time evolution for the double well potential.  
  Parameters: $\mu^2=1$, $\lambda=3.0$, $\phi(0)=0.4$,
  $T_0=0.1$ (a) time evolution of the mean field
  $\phi(t)$; (b) time evolution of the field $\chi(t)$; the solid lines 
  relate to the two-loop
  2PPI approximation, the dashed lines to the one-loop 2PPI or Hartree
  approximation, the dotted lines to the 2PI-1/N approximation and
  the dashed dotted
  lines to the BVA approximation, where the last two simulations are taken
  from \cite{Cooper:2002qd}}
\end{figure}

\begin{figure}[htbp]
  \centering
(a)\hspace{-0.5cm}\includegraphics[width=8cm]{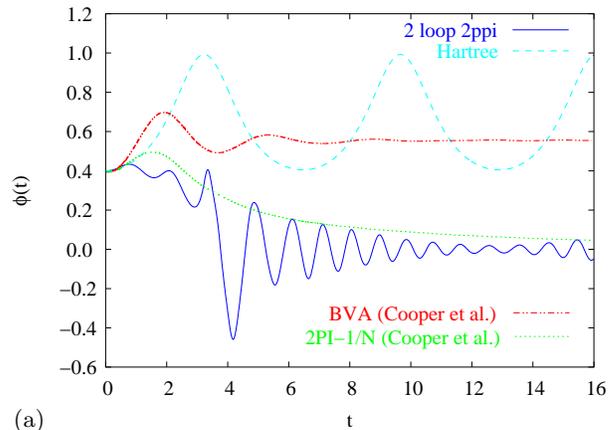}
\par \vspace{0.5cm}
%\qquad
(b)\hspace{-0.5cm}\includegraphics[width=8cm]{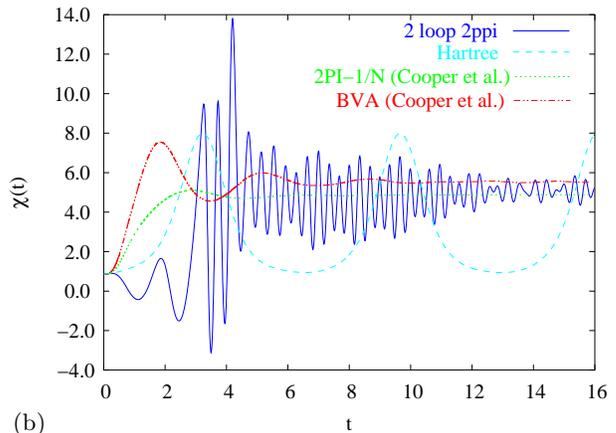}
  \caption{\label{fig:phi-l21.3} Same as Fig.~\ref{fig:phi-l3.0} but
    for $\lambda=21.3$} 
\end{figure}

\begin{figure}[htbp]
  \centering
\includegraphics[width=8cm]{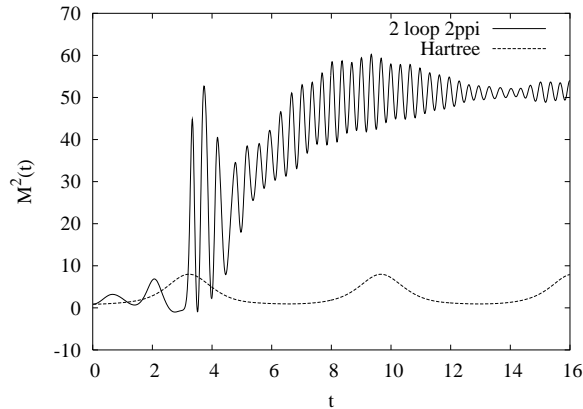}
  \caption{\label{fig:calm-l21.3} Time evolution of the effective
    mass $\calm^2(t)$ for the simulation in Fig.~\ref{fig:phi-l21.3}; the
    solid line represents the two-loop 2PPI approximation, the dashed
    line the one-loop or Hartree approximation} 
\end{figure}

\begin{figure}[htbp]
  \centering
\includegraphics[width=8cm]{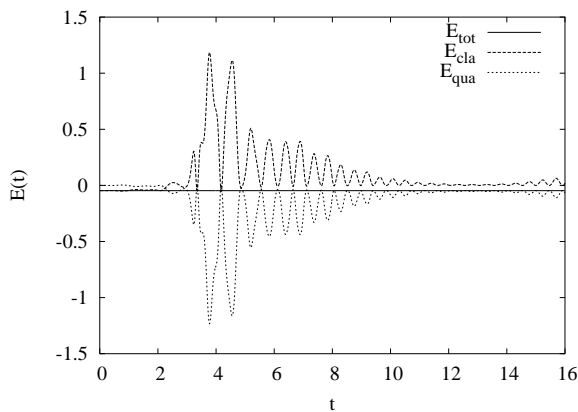}
  \caption{\label{fig:energy-l21.3} Time evolution of the energy for the
    simulation in Fig.~\ref{fig:phi-l21.3}; the solid line is the
    total energy, the dashed line the classical energy and the dotted
    line the quantum energy} 
\end{figure}

The time evolution of the mean field $\phi(t)$ is presented in 
Fig.~\ref{fig:phi-l3.0}a and Fig.~\ref{fig:phi-l21.3}a and the time evolution 
of $\chi(t)$ is displayed in Fig.~\ref{fig:phi-l3.0}b 
and Fig.~\ref{fig:phi-l21.3}b. These are compared with the
results obtained in the Hartree approximation, and to those
obtained by Cooper et al. for the BVA and the 2PI-1/N approximations.
For the simulation with $\lambda=21.3$ we 
additionally present the time evolution of the effective mass $\calm^2(t)$ in 
Fig.~\ref{fig:calm-l21.3} and the energy contributions in 
Fig.~\ref{fig:energy-l21.3}.

Surprisingly the mean field obtained in the 2PPI approximation
takes off rather early from the results obtained in the other
approximations. It  starts to oscillate around
$\phi=0$ for $t\gtrsim 4$, this indicates the absence of spontaneous
symmetry breaking. The amplitude of the oscillations decreases with time. 
Remarkably, the mean field of the 2PI-1/N approximation decreases with
the same time scale and similar amplitude, but without oscillations, 
forming a kind of envelope.

The composite field $\chi(t)$ drives to negative values at early times 
and oscillates around 
positive values for later times [Figs.~\ref{fig:phi-l3.0}b 
and~\ref{fig:phi-l21.3}b]. For the simulation with $\lambda=3.0$ 
the transition between the two regimes sets in at a time $t\gtrsim 13$, 
whereas for the simulation with $\lambda=21.3$ 
it takes place for $t\gtrsim 4$. While in the simulation with 
$\lambda=21.3$ the average value of the field $\chi(t)$ in the two-loop 2PPI 
approximation is, for later times, very close to the BVA and 2PI-1/N results, 
the situation for $\lambda=3.0$ is manifestly different. Here the average 
value of $\chi(t)$ in the two-loop 2PPI approximation lies significantly 
below the one obtained in the other approximations. 
As we have mentioned in the previous section the comparison is not 
on firm grounds.
 
Both the mean field $\phi$ and the field $\chi$ in the two-loop 2PPI 
approximation oscillate more strongly than the ones of the
BVA and  2PI-1/N approximations. However the amplitude
is much smaller than the one of the Hartree approximation,
and it decreases with time.

In Fig.~\ref{fig:calm-l21.3} we display the 
time evolution of the variational mass $\calm^2(t)$ in the two-loop 2PPI 
approximation for the simulation with $\lambda=21.3$. As the difference 
of $\chi(t)$ and $\calm^2(t)$ is directly proportional to  
$\Delta^{(2)}(t)$ [see Eq.~\ref{eq:chidef}], one concludes that a large 
contribution in $\calm^2(t)$ comes from the two-loop term 
[Fig.~\ref{fig:vacgraphs}b]. The energy displays strong oscillations
at the transition towards the symmetric phase. At late times
the energy has been transferred to the quanta almost entirely.

\section{\label{sec:conclusions}Summary, Conclusions and Outlook}

We have compared numerical simulations in 
the two-loop 2PPI approximation to the Hartree approximation 
and to results in the BVA and 2PI-1/N approximation by Cooper et al..

In contrast to the Hartree and BVA approximations the two-loop 2PPI 
approximation exhibits no spontaneous symmetry breaking 
in the nonequilibrium evolution of $\Phi^4$ quantum field theory. The 
mean field $\phi(t)$ tends to the symmetric configuration with $\phi=0$. 
The same is true for the 2PI-1/N approximation. 
It is somewhat surprising
that an approximation which is less powerful in resumming higher 
order graphs fares well in reproducing the expected phase structure. 
We think that this deserves further investigation.

In $3+1$ dimensions the two-loop 2PPI approximation has been
applied \cite{Smet:2001un, Baacke:2002pi} to the O(1) 
and O(N) models in thermal equilibrium. There 
the approximation displays a second-order
phase transition as expected for the exact theory.
It therefore seems promising to study the out-of-equilibrium
evolution in these models as well. Work on this is in progress.

We thank Stefan Michalski for helpful conversations.

\end{document}